\documentclass[twocolumn]{aastex631}

\usepackage{graphicx}
\usepackage{float}

\usepackage{subfigure}

\begin{document}

\title{3D-DASH: The evolution of size, shape, and intrinsic scatter in populations of young and old quiescent galaxies at $0.5<z<3$}

\author[0009-0001-4005-5490]{Maike Clausen}
\affiliation{Max Planck Institute of Astronomy, Königstuhl 17, 69117 Heidelberg,Germany}

\author[0000-0001-7160-3632]{Katherine E. Whitaker}
\affiliation{Department of Astronomy, University of Massachusetts, Amherst, MA 01003, USA}
\affiliation{Cosmic Dawn Center (DAWN), Denmark}

\author[0000-0003-1665-2073]{Ivelina Momcheva}
\affiliation{Max Planck Institute of Astronomy, Königstuhl 17, 69117 Heidelberg,Germany}

\author[0000-0002-7031-2865]{Sam E. Cutler}
\affiliation{Department of Astronomy, University of Massachusetts, Amherst, MA 01003, USA}

\author[0000-0002-1714-1905]{Katherine A. Suess}
\affiliation{Department of Astronomy and Astrophysics, University of California, Santa Cruz, 1156 High Street, Santa Cruz, CA 95064 USA}
\affiliation{Kavli Institute for Particle Astrophysics and Cosmology and Department of Physics, Stanford University, Stanford, CA 94305, USA}

\author[0000-0003-1614-196X]{John R. Weaver}
\affiliation{Department of Astronomy, University of Massachusetts, Amherst, MA 01003, USA}

\author[0000-0001-8367-6265]{Tim Miller}
\affiliation{Department of Astronomy, Yale University, 52 Hillhouse Ave., New Haven, CT, 06511, USA}

\author[0000-0002-5027-0135]{Arjen van der Wel}
\affiliation{Ghent University, Krijgslaan 281, Building S9, 9000 Ghent, Belgium}

\author[0000-0003-3735-1931]{Stijn Wuyts}
\affiliation{Department of Physics, University of Bath, Claverton Down, Bath, BA2 7AY, UK}

\author[0000-0002-6047-1010]{David Wake}
\affiliation{Department of Physics and Astronomy, 107 Rhoades/Robinson Hall, University of North Carolina at Asheville}

\author[0000-0002-8282-9888]{Pieter van Dokkum}
\affiliation{Astronomy Department Yale University, 52 Hillhouse Ave, New Haven, CT 06511}

\author[0000-0001-5063-8254]{Rachel S. Bezanson}
\affiliation{Department of Physics and Astronomy and PITT PACC, University of Pittsburgh, Pittsburgh, PA, USA}

\author[0000-0003-2680-005X]{Gabriel Brammer}
\affiliation{The Cosmic Dawn Center, Jagtvej 155A, 2200 København N}

\author[0000-0002-8871-3026]{Marijn Franx}
\affiliation{Leiden Observatory, P.O. Box 9513, 2300 RA, Leiden, The Netherlands}

\author[0000-0002-7524-374X]{Erica J. Nelson}
\affiliation{Department for Astrophysical and Planetary Science, University of Colorado, Boulder, CO 80309, USA}

\author[0000-0003-4264-3381]{Natasha M. Förster Schreiber}
\affiliation{Max-Planck-Institut für extraterrestrische Physik
Giessenbachstr. 1, D-85748 Garching, Germany }

\begin{abstract}

We present a study of the growth of the quiescent galaxy population between $0.5<z<3$ by tracing the number density and structural evolution of a sample of 4518 old and 583 young quiescent galaxies with log(M$_{\star}$/M$_{\odot}$)$>$10.4, selected from the COSMOS2020 catalog with complementary HST/F160W imaging from the 3D-DASH survey.
Among the quiescent population at $z$$\sim$2, roughly 50\% are recently quenched galaxies; these young quiescent galaxies become increasingly rare towards lower redshift, supporting the idea that the peak epoch of massive galaxy quenching occurred at $z$$>$2.
Our data show that while the effective half-light radii of quiescent galaxies generally increases with time, young quiescent galaxies are significantly smaller than their older counterparts at the same redshift.
In this work we investigate the connection between this size difference and other structural properties, including axis-ratios, color gradients, stellar mass, and the intrinsic scatter in effective radii.
We demonstrate that the size difference is driven by the most massive sub-population (log(M$_{\star}$/M$_{\odot}$)$>$11) and does not persist when restricting the sample to intermediate mass galaxies (10.4$<$log(M$_{\star}$/M$_{\odot})$$<$11).
Interestingly, the intrinsic scatter in physical size shows a strong co-evolution over the investigated time period and peaks around z$\sim$2 for both populations, only diverging at $z<1$.
Taken together, and assuming we are not missing a significant population of lower surface brightness galaxies, while the formation and quenching mechanisms that dominate at higher redshifts yield compact remnants, multiple evolutionary pathways  may explain the diverse morphologies of galaxies that quench at $z$$<$1.

\end{abstract}

\section{introduction}

The structural evolution and growth of quiescent galaxies over cosmic time is well established \citep[e.g.,][]{cimatti2008,vandokkum2008,damjanov2009,vandersande2013,patel2017,mowla2019,mathaduru2020} but the physical processes causing this growth are still subject to ongoing research. 
Theories attribute the growth in the average size of red-sequence galaxies to a combination of minor mergers and accretion that puff up individual passively evolving galaxies over time \citep[e.g.,][]{bezanson2009,naab2009}, in combination with growth of the population on average due to larger sizes of the galaxies being quenched at later times \citep[progenitor bias; e.g.,][]{vandokkum2008,vanderwel2009,szomoru2011,Carollo2013}. 
The former has been demonstrated empirically through the gradual build up of the average stellar mass profiles \citep{vandokkum2010,hill2016} and direct studies of minor merger rates \citep[e.g.,][]{newman12,belli15,suess23}. 
In the latter process, the most recently quenched galaxies form later and are thus larger because the star-forming galaxies from which they form had more time to grow \citep[e.g.,][]{Poggianti2013b}.

The most efficient approach to test whether minor merger growth or progenitor growth is more important in driving size evolution is to study the most recently quenched galaxies in comparison with older quiescent galaxies: progenitor bias predicts that young quiescent galaxies should be larger than older quiescent galaxies at similar epochs \citep[][]{Ichiwaka2017,pawlik2019,Wu2020}.
However, there is an ongoing discussion as to whether or not progenitor bias alone can explain the size evolution at higher redshifts.
Some studies find that recently quenched galaxies are instead either similar in size or more compact than older galaxies at $z$$>$1 
\citep[e.g.,][]{whitaker2012,almaini2017,chen2022}.

There exists a major shortcoming in these earlier investigations: robustly quantifying the sizes of high redshift quiescent galaxies with ground-based near-infrared imaging is challenging due to their compact nature. 
Moreover, the situation is further complicated by ``outshining": young stars dominate the integrated light from galaxies, outshining by factors of 10 or more per unit mass of the low-luminosity old stars \citep[e.g.,][]{papovich2001}.  
\citet{Suess2020} show that accounting for the outshining effect can vastly reduce or even remove the offset in the sizes of young and old quiescent galaxies. 

On the other hand, if this size discrepancy is real, the physical processes driving the quenching of early galaxies may be different from those that dominate at later times \citep[e.g.,]  []{wild2016,maltby2018,belli2019,Wu2020,tacchella2022}.
At early times, galaxies undergo a period of quick star-formation either through the dissipational collapse of gas and dust clouds or by rapid mergers of smaller galaxies accompanied by a brief quenching phase. 
This scenario predicts smaller, compact quiescent galaxies at high redshift \citep{williams2009,wellons2015}.
At later times larger, gas-rich, star-forming galaxies appear to quench slowly, with a variety of physical processes resulting in larger (less compact) quiescent galaxies \citep[e.g.,][]{peng2010,wild2016,park2022}.
While this could, in theory, explain the redshift dependent trends, these studies are based on effective half-light radii, meaning they rely on the radius including 50\% of the light of a galaxy as a size measurement.
However, as previously mentioned, light is a biased tracer of galaxy size since it does not necessarily trace mass consistently \citep[e.g.,][]{suess2019b,mosleh20,Suess2022,miller2022}.
Therefore, the established light-weighted studies of size evolution of quiescent galaxies need to be re-evaluated using higher resolution near-infrared (NIR) imaging.

In this work, we use the high resolution Hubble Space Telescope (HST) NIR imaging from the 3D-DASH survey \citep{Mowla2022} to compare the morphologies of an unprecedentedly large sample of young and old quiescent galaxies between the redshifts $0.5<z<3$ in the COSMOS extragalactic field. 
By covering 1.7 square degrees, we are well positioned to reassess the results from earlier ground-based studies in a meaningful way, as well as to bridge the high redshift and low redshift studies. 
We discuss our sample selection in Section \ref{sec:sampleselection} and measure the evolution of the number density in Section \ref{sec:numberdensity}.
Section \ref{sec:sizeevolution} combines studies of the redshift evolution of size, axis-ratio, intrinsic scatter in size as well as impact of stellar mass for young versus old quiescent galaxies.
In this section, we augment our analysis with independent half-mass radius measurements from \cite{suess2019b,suess2019a} and \citet{miller2023} to compare half-mass to half-light ratios for our selection criteria to test the observed size evolution for outshining effects.
We close with the discussion in Section \ref{sec:discussion} and summarize our conclusions in Section \ref{sec:conclusion}.
 
Throughout this work we assume a $\Lambda$CDM cosmology with $\Omega_M = 0.3$, $\Omega_{\Lambda}=0.7$ and $H_0 = 70$ km s$^{-1}$ Mpc$^{-1}$, as well as a \citet{chabrier2003IMF} initial mass function for stellar masses.

\section{Data and Sample selection} \label{sec:sampleselection}

In this work we combine data from the COSMOS2020 catalog \citep{weaver2022} with the 3D-DASH morphological catalog \citep[see methodology in][]{Cutler2022}, based on the $H_{\mathrm{F160W}}$ image released in \citet{Mowla2022}.

The COSMOS2020 catalog combines observations  from {\it GALEX} \citep{zamojski2007}, ACS/HST \citep{leauthaud2007}, Subaru-HSC \citep{aihara2019subarurelease}, VISTA \citep{mccracken2012Uvistarelease,moneti2019Ultravistarelease}, Spitzer/IRAC \citep{ashby2013,ashby2014,ashby2015,ashby2018} and  CFHT/MegaCam/CLAUDS \citep{sawicki2019CLAUDSrelease}, covering 2 deg$^2$ of the COSMOS field.
The catalog uses 17 filters from the FUV to IR and a new profile-fitting photometric extraction tool \textsc{The Farmer} \citep{weaver2022} to measure the properties of $\approx 964,500$ sources providing the measurements for sSFR, stellar masses and colors used in this work.
For our selection, we use only objects measured in all bands that are outside of bright stars and known artifacts (\texttt{FLAG\_COMBINED}) resulting in 1.27 deg$^2 $ of coverage, as well as the recommended threshold of $K_s <25.0$ ABmag.
To ensure selection of only well-characterized sources, we also require $\chi^2_{\rm galaxy}/\chi^2_{\nu} < 20$.

The 3D-DASH morphological catalog is based on the imaging portion of the 3D-DASH survey in the $H_{160}$ filter \citep{Mowla2022}.
It consists of 1256 WFC3 pointings carried out with the `Drift-and-Shift' (DASH) technique for HST/WFC3 imaging combined with CANDELS/3D-HST archival data to cover a total of 1.43 deg$^2$ in the COSMOS field.
We use GALFIT \citep{peng2002} to determine the morphological properties, including effective half-radii and circular radii $\sqrt{b/a} \times r_e$.
For our sample selection, we use the recommended flags for deblending \citep{weaver2022} and to ensure robust GALFIT measurements \citep{vanderwel2012,Cutler2022}.

\begin{figure}
    \centering
    \includegraphics[width=\columnwidth]{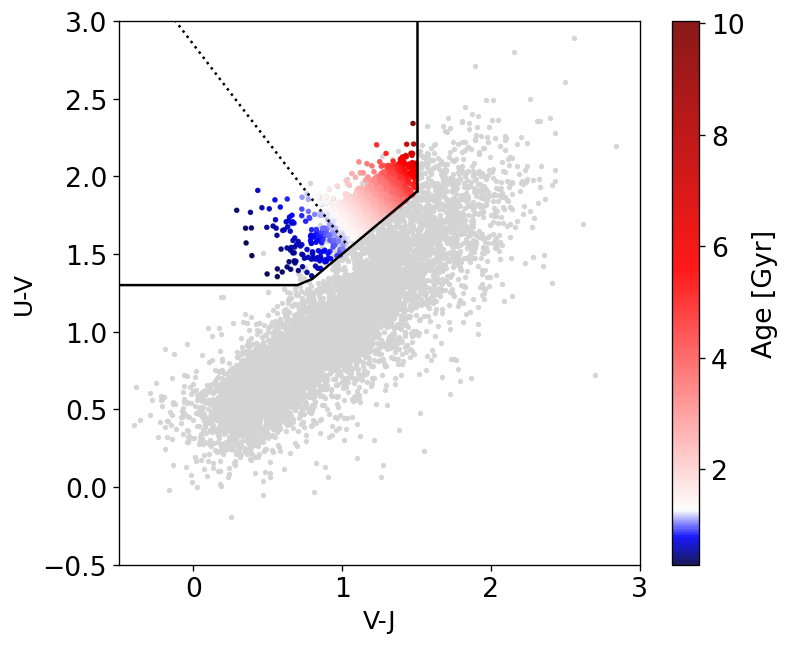}
    \caption{UVJ selection of quiescent galaxies at $0.5<z<3$, color-coded according to stellar ages predicted from the age-color trend in \citet{belli2019}.The full sample of galaxies from the COSMOS2020 survey is shown in light gray. We divide our sample into 4518 old and 583 young quiescent galaxies (closely matching earlier work by \cite{whitaker2013}, dashed line), corresponding to stellar ages of older/younger than $\sim$ 1 Gyr.}
    \label{fig:uvj-selection}
\end{figure}
\begin{figure}
    \centering
    \includegraphics[width=\columnwidth]{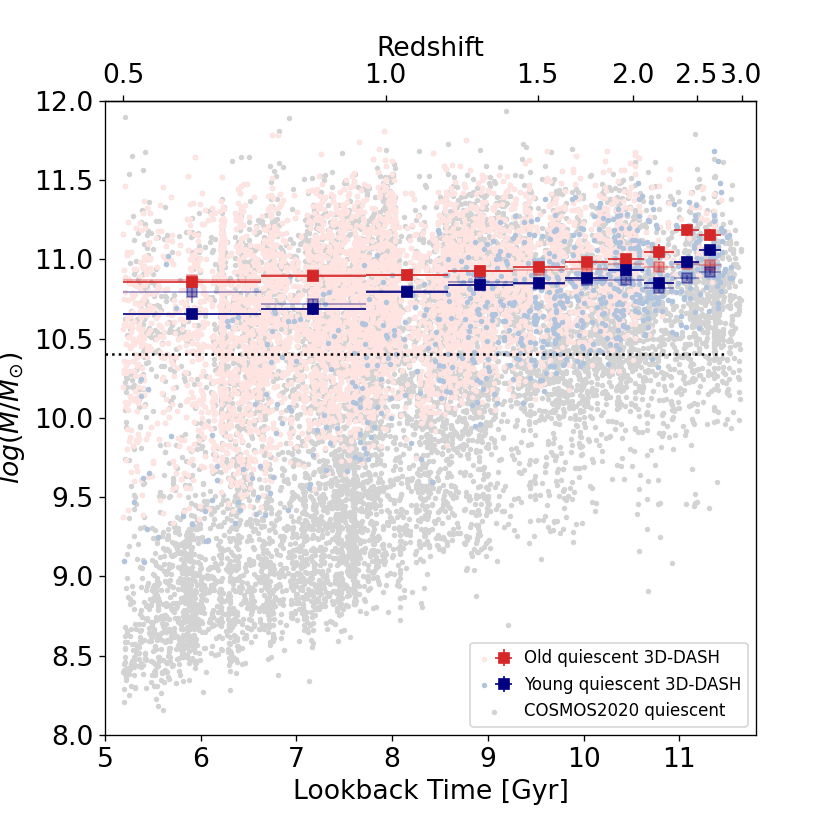}
    \caption{Total (light gray) and selected sample of young (bright blue) and old (bright red) quiescent galaxies depending on mass and redshift. For our analysis we only look at massive quiescent galaxies with log(M$_{\star}$/M$_\odot$)$>$10.4 (black dotted line). We include the average mass of galaxies above that limit in red (old) and blue (young).}
    \label{fig:redshicft_mass}
\end{figure}

Additionally, we require that all sources in our sample are well detected in both the 3D-DASH imaging and COSMOS2020 K-band detection image (median SNRs of 825.93 for COSMOS2020 and 10.01 for 3D-DASH, respectively).  
We merge the two catalogs by cross-matching the coordinates within a 0.2\arcsec \;radius.
We select a stellar mass threshold of M$_* >$10$^{10.4}$M$_{\odot}$.
Of the sample, 99.78\% are brighter than the applied $K_s <25$ ABmag cut in COSMOS2020 and satisfied this stellar masscut and 97.94\% are brighter than the  $H <23.4$ ABmag cut in 3D-DASH.
77.75\% of galaxies above the stellar mass threshold have a robust GALFIT measurement following the recommendations set forth in \citet{Cutler2022} specific to the DASH imaging.
Combining all criteria, the final sample contains 76.57\% of galaxies above our chosen mass limit.

Figure \ref{fig:redshicft_mass} shows that our sample starts to be incomplete at $z>2$ due to the fainter nature of the 3D-DASH survey which does not allow precise morphology measurements at these high redshifts and lower masses.
We will show in Section \ref{sec:discussion} that this incompleteness does not affect our results.

Young quiescent galaxies, with stellar ages of approximately 1 Gyr or less, are therefore easily identified as A-type stars and dominate the rest-frame optical light in their spectral energy distributions (SEDs).   
Following \citet{whitaker2013} we use a rest-frame color selection box of $(U-V)> 0.8(V-J)$, $(U-V)<1.3$ and $(V-J) <1.5$ as a criterion for quiescence, as shown in Figure \ref{fig:uvj-selection} \citep[see also][]{Labbe2005,wuyts2007,williams2009}. 
Despite the popularity of the UVJ selection method at $z<3$, some recently published works question the purity of such a sample at higher redshifts, arguing that highly dust obscured star-forming galaxies enter the area of old quiescent galaxies in the UVJ diagram \citep[e.g.][]{antwi-danso2022,gould2023}. 
In the redshift regime explored in this paper, typical failure rates of 20\% are reported from observations \citep{schreiber2018,leja2019} and as high as 30\% in simulations \citep[e.g.][]{lustig2022,akins2022}.

Following \citet{leja2019}, we test the purity of our sample by comparing the fraction of quiescent galaxies identified via UVJ selection to a sSFR selection (as derived in the COSMOS2020 catalog) and conclude that the purity of the UVJ selection is $>85\%$ for a threshold of $\mathrm{\log(sSFR})<-11$ and $>95\%$ for $\mathrm{\log(sSFR})<-9.5$ \citep[e.g.,][]{brinchmann2004,fontanot2009}.

Thus, we acknowledge that there may be some minimal contamination inherent to our quiescent classification scheme, but note that the vast majority of our sample is securely quiescent.
We consider our sample sufficiently pure so as to not affect the validity of our analysis.

There is a tight relationship between stellar age of quiescent galaxies and their rest-frame colors within the UVJ diagram that we can use to distinguish between old and young galaxies \citep[e.g.,][]{whitaker2012,whitaker2013,belli2019}.
First proposed in \citet{whitaker2012} based on photometry alone, this trend was later validated with grism spectroscopy in \citet{whitaker2013} and high-resolution, deep ground-based spectroscopy in \citet{belli2019}.  
Within the UVJ quiescent region, younger (stellar age $<$1 Gyr) and older (stellar age $>$1 Gyr) quiescent galaxies occupy distinct regions.  
In this work, we adopt the best-fit relation published in \citet{belli2019} and divide the sample at 1 Gyr; this selection closely follows the original division proposed in \citet{whitaker2013} (see dotted line, $(U-V) < -1.25 (V-J) + 2.58$, in Figure~\ref{fig:uvj-selection}).  
The predicted ages closely align with ages published based on the COSMOS2020 photometry in \citet{weaver2022}. 

All together, we select massive galaxies (log(M$_{\star}$/M$_\odot$)$>$10.4) and restrict our selection to the redshifts between $0.5<z<3$, dividing our sample into 4518 old and 583 young quiescent galaxies.  
We show our selection based on the UVJ diagram in Figure \ref{fig:uvj-selection}, color-coded by the predicted age from the \citet{belli2019} age-color trend, relative to the COSMOS2020 parent sample in light gray.
The noted variability in galaxy distributions across redshift is attributable to the well-documented large-scale structures within the COSMOS field \citep[e.g.,]{gilli2003, vanzella2005}; however, this variability does not significantly alter the outcomes of this study due to the employed bin sizes.
Figure \ref{fig:redshicft_mass} shows the mass-redshift evolution of the selected old (red) and young (blue) quiescent galaxies relative to the galaxies in the COSMOS2020 parent sample (light gray).

\section{Number Density Evolution} \label{sec:numberdensity}
Before digging into the structural evolution, we first need to gain a better understanding of how the populations of young and old quiescent galaxies build up in number over time.  

\begin{figure}
    \centering
    \includegraphics[width=\columnwidth]{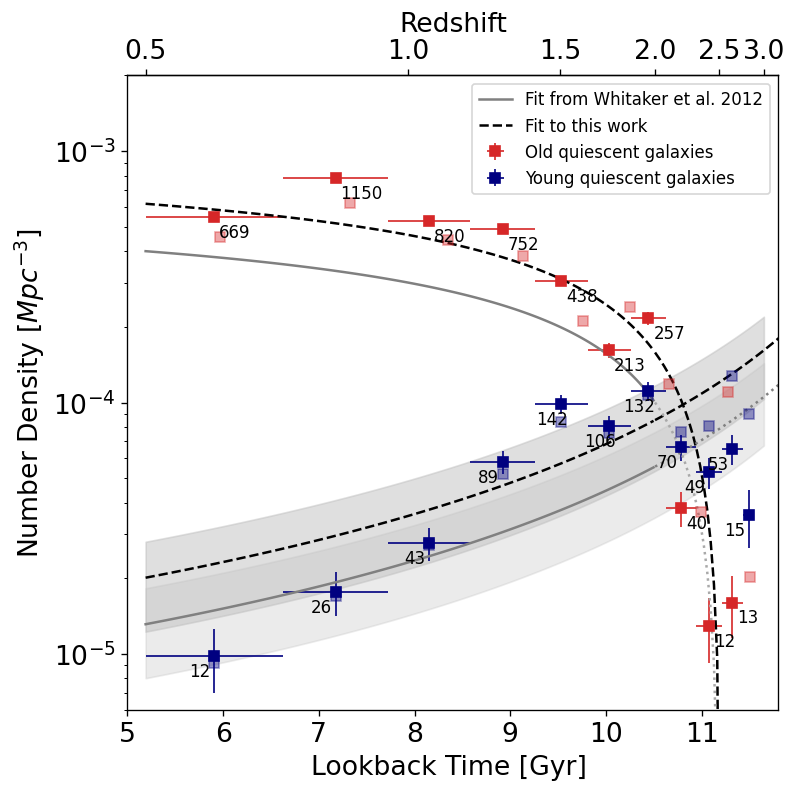}
    \caption{The number density evolution of young (blue) and old (red) quiescent galaxies reveals different behaviors over cosmic time. Old quiescent galaxies decrease in number density towards higher redshift while young quiescent galaxies initially increase and then follow the old quiescent galaxies evolution after their peak at around $z\approx2$. The number of galaxies in every bin is shown. The errors along the x-axis correspond to the bin in redshift in which the galaxies are counted; while the errors along the y-axis are computed using a combination of Poisson statistics and relative cosmic variance following \citet{Somerville04}. The lighter squares depict the number densities of the two samples derived from the deeper COSMOS2020 catalog. We include the original fit from \citet{whitaker2012} and use their equation to fit to our data.}
    \label{fig:numberdensity}
\end{figure}
\begin{figure*}
    \centering
    \includegraphics[width=\textwidth]{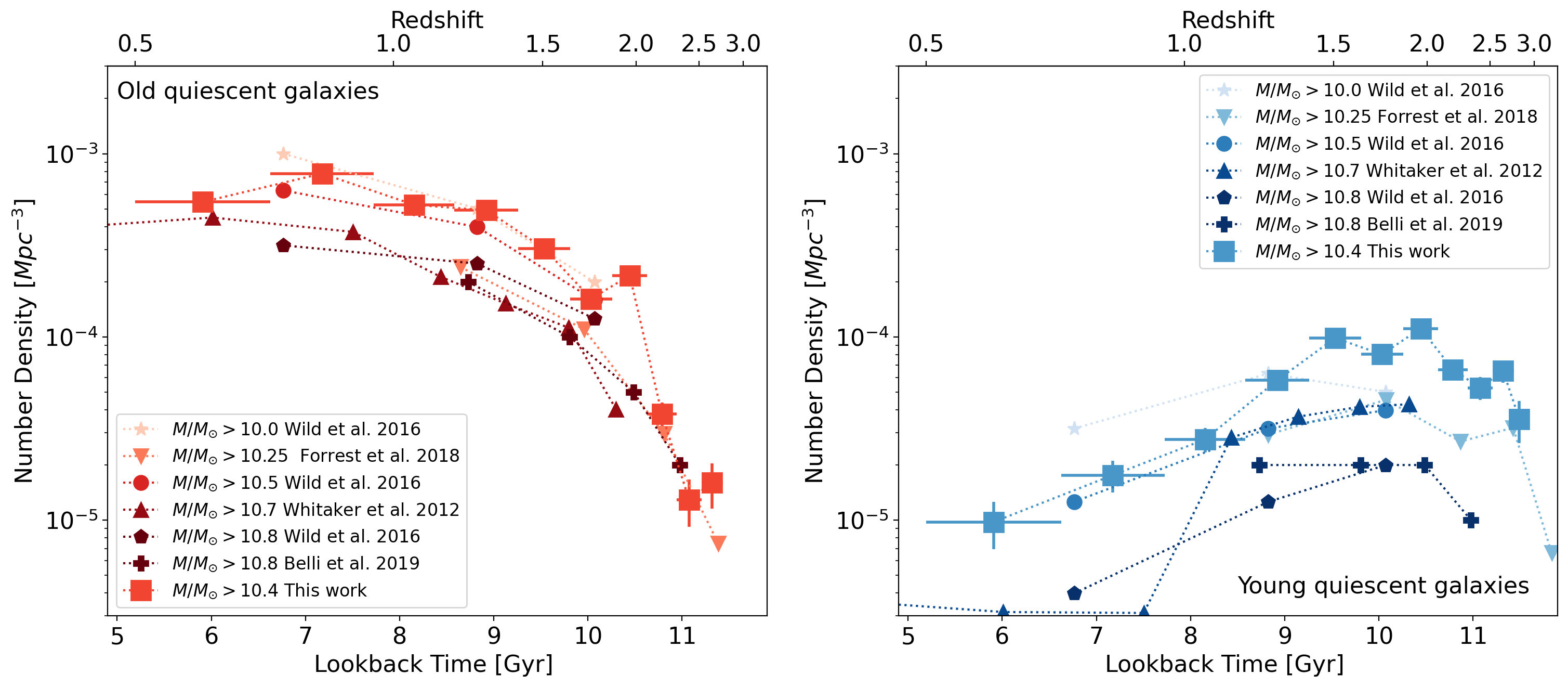}
    \caption{Number densities from \cite{belli2019,wild2016,forrest2018,whitaker2012} of old (left panel) and young quiescent/PSB (right panel) galaxies with varying thresholds in stellar mass. A higher stellar mass limit reduces the number density, thereby decreasing the normalization, but does not change the shape of the overall evolution.}
    \label{fig:numberdensity_comparisson}
\end{figure*} 

Figure \ref{fig:numberdensity} shows the number density evolution of our sample of old (red) relative to young (blue) quiescent galaxies out to $z$=3. 
In order to calculate the number density, we divide the redshift range $0.5<z<3$ into separate bins and divide the number of galaxies within each bin by the corresponding co-moving volume.
The errors in number density are determined by adding the Poisson error and the cosmic variance error calculated according to \cite{Somerville04} in quadrature.
The errorbars on the x-axis depict the size of the redshift bin within which the corresponding number density is computed. 
For context, we label the number of galaxies in each bin next to the symbols. 

Figure \ref{fig:numberdensity} illustrates that the number density for old quiescent galaxies increases towards lower redshift and plateaus below $z\approx 1.2$.
At $z<2.5$, our measurements concur well with number densities derived using just the COSMOS2020 catalog \citep{weaver2022b} which are shown in fainter squared symbols without errorbars.

Following \citet{whitaker2012}, we fit the function $a-b\times(1+z)$ to the measured number density evolution resulting in the fit-parameters $a=0.0013$ and $b=0.0004$.
Apart from one outlier between $0.7<z<0.9$ this function describes our data well.
This outlier possibly stems from known clusters in the COSMOS field at $z\approx0.7$ and $z\approx 0.9$ driving the number density of old quiescent galaxies unusually high \citep[e.g.][]{gilli2003,vanzella2005}.

Contrary to the 3D-DASH sub-sample, the number densities of young quiescent galaxies in the COSMOS2020 parent sample continue to increase at $z>2.5$.
This deviation is a result of the shallower nature of the 3D-DASH imaging, which is limiting detection at these redshifts (see Fig. \ref{fig:redshicft_mass}).
We explore the effects of this selection in Section \ref{sec:discussion}.

The number density evolution for young quiescent galaxies differs significantly from older quiescent galaxies at lower redshift.
At $z<1$ young quiescent galaxies appear to be rare, with number densities almost two orders of magnitude lower than for old quiescent galaxies \citep[see also][]{whitaker2012}.
The number density of young quiescent galaxies increases towards higher redshifts until the population peaks around $z \approx 2$, after which the 3D-DASH sub-population appears to decrease in number density compared to the old population.

We note that the DASH sample begins to become incomplete at $z\gtrsim2$, which could cause this apparent drop in the number density of young quiescent galaxies at high redshift.
To test this, we repeat the same analysis with the parent COSMOS2020 catalog which indicates constant or even increasing number densities for young quiescent galaxies towards higher redshift. 
The latter would be consistent with the findings of \citet{gould2023}.

The number density of young quiescent galaxies is assumed to be closely connected to the old population through 
\begin{equation}\label{eq:N_y}
    N_y(t) = N_o(t+\tau)+N_o(t)
\end{equation}
where $t$ is the age of the universe and $N_y$/$N_o$ describe the number density of young/old quiescent galaxies respectively, following \citet{whitaker2012}.
We assume the timescale for quenching to be $\tau=0.5$ Gyr and vary $\tau$ to $0.3$ Gyrs to include the uncertainty in this assumption (gray area in Figure \ref{fig:numberdensity}). 
The basic premise of this simple model is that galaxies must first pass through the young quiescent phase before aging into the older population. 
This model roughly describes the data at $z<2$ thereby validating both the assumption of the linear evolution of the number density of old quiescent galaxies as well as the assumed model for young quiescent galaxies (see Equation \ref{eq:N_y}).
We further include the fits from \citet{whitaker2012} ($0.0007-0.0002\times(1+z)$) in Figure \ref{fig:numberdensity}.  
\cite{whitaker2012} apply a higher mass limit to their sample (log(M$_{\star}$/M$_\odot$)$>$10.8), explaining the systematic offset of their measurements towards lower number densities given the shape of the galaxy mass function \citep[e.g.,][]{Marchesini2009,leja2020}.

To further study the normalization of the number density evolution, we compare the overall trends observed by previous studies of old versus young quiescent galaxies in the literature  \citep{whitaker2012,wild2016,forrest2018,belli2019}, color-coding by the applied stellar mass thresholds in Figure \ref{fig:numberdensity_comparisson}.
These earlier studies probe smaller volumes, which will significantly increase cosmic variance and therefore the uncertainties of individual values of number density, especially for the young quiescent population.
Despite this limitation, the left panel of Figure \ref{fig:numberdensity_comparisson} demonstrates how all measurements of old quiescent galaxies follow the same evolution, with more massive samples showing lower number densities below $z$=2. 
This is expected in the \citet{Thomas2015} downsizing paradigm where more massive galaxies form and quench earlier.
The right panel of Figure \ref{fig:numberdensity_comparisson} displays the same general trend for young quiescent galaxies and validates the measured evolution, only differing in magnitude due to the different applied lower mass limits.
Both trends qualitatively and quantitatively agree with previous measurements.

\section{Structural Evolution} \label{sec:sizeevolution}
We next turn towards examining the morphological evolution with redshift of the two quiescent sub-populations, focusing on the redshift evolution of size (Section~\ref{subsec:size}) and axis ratio (Section~\ref{subsec:axratio}), the impact of color gradients  (Section~\ref{subsec:colorgradients}) and stellar mass (Section~\ref{subsec:massdependence}), as well as the intrinsic scatter of both sub-populations (Section~\ref{subsec:scatter}).

\subsection{Size Evolution}
\label{subsec:size}

\begin{figure*}
    \centering
    \includegraphics[width=\textwidth]{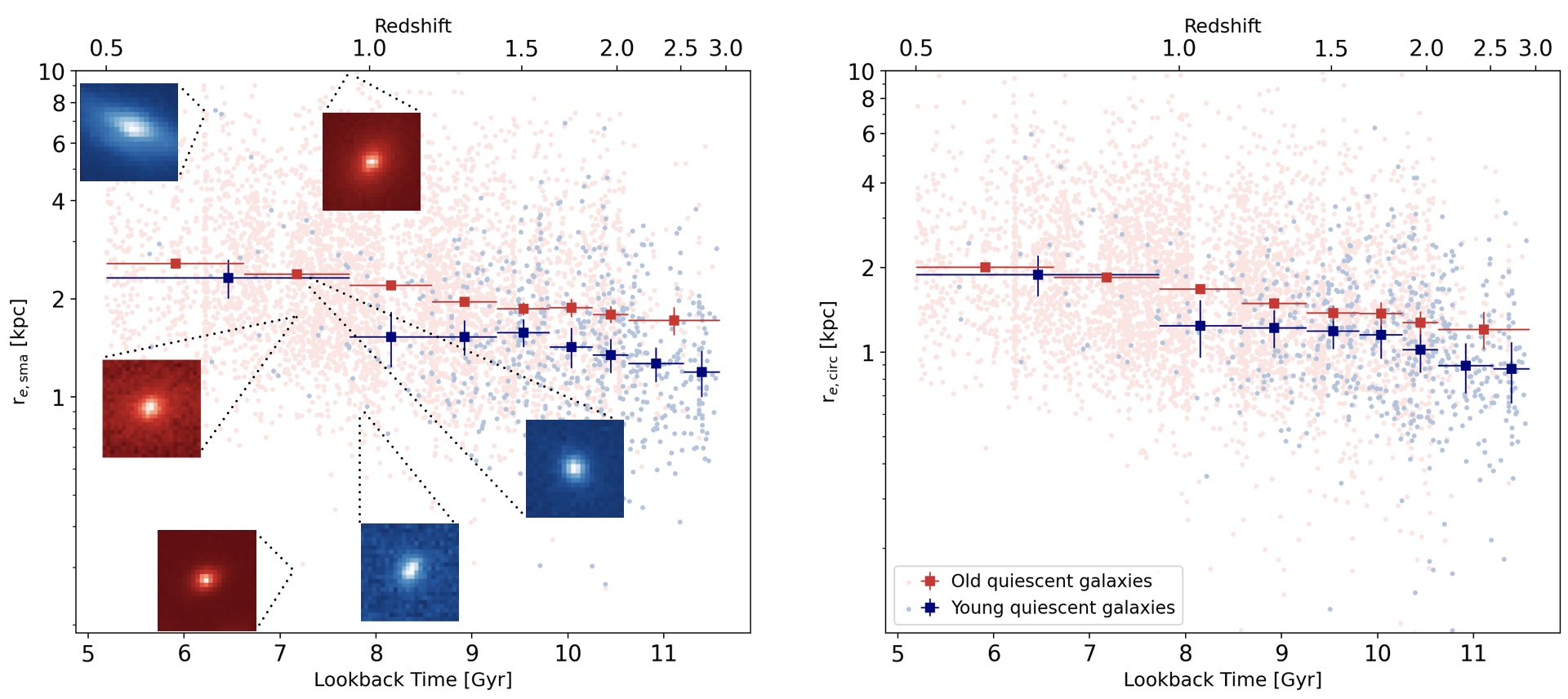}
    \caption{\textbf{Left panel:} There is a strong evolution of effective half-light radii of the old (light red) and young (light blue) quiescent galaxies as a function of lookback time. Young quiescent galaxies are on average smaller than old ones at higher redshift $z>1$ while they grow with the old sample and demonstrate similar average sizes at low redshift. \textbf{Right panel:} Circularized effective half-light radii of the two populations. Both populations replicate the evolution of the left panel of Figure \ref{fig:sizeevolution} although the offset is less significant when considering circularized radii.}
    \label{fig:sizeevolution}
\end{figure*}

The left panel of Figure \ref{fig:sizeevolution} shows the evolution of major-axis half-light radii of both old (red) and young (blue) quiescent galaxies. 
The mean size of both populations increases towards lower redshift. 
In Figure \ref{fig:sizeevolution}, the square symbols represent the geometric means of the effective half-light radii of both populations while the individual measurements are shown as small transparent circles, color-coded according to their sub-population classification.
To prevent uncertainties resulting from low number statistics, we merge adjacent redshift bins with fewer than 20 galaxies for the remainder of this paper.
The error bars along the y-axis represent the error of the mean of the distribution while the errors along the x-axis span the range of redshift over which the average is calculated.

The increase in size of quiescent galaxies globally has been well studied \citep[e.g,][]{mowla2019,patel2017,mathaduru2020} and is in agreement with recent observations of galaxies in general \citep[e.g,][]{shibuya2015,yano2016}. 
The size evolution is more pronounced for the young quiescent sample.
At intermediate redshifts ($z$$\sim$0.5-1), the samples have the same average size; at $z$$>$1, young quiescent galaxies are significantly ($\approx3\sigma$) smaller than old quiescent galaxies. 

Some studies use the circularized radius ($r_{e,circ} = r_{e,sma} * \sqrt{b/a}$) as an alternative measurement of size.
To ensure a comprehensive view on size evolution, we therefore demonstrate in Figure \ref{fig:sizeevolution} that the trends in semi-major axis half-light effective radius are reproduced when using circularized radii:
both populations increase in size on average towards lower redshift, with the young quiescent galaxies being smaller than the old populations at higher redshift.
However, it should be noted that the difference in size is less significant with circularized radii as size-measurement.
This alternative method of measuring sizes is more susceptible to viewing angle affects than the semi-major axes measurement \citep{vandeven2021}.
A flattened axisymmetric disk system viewed from a higher inclination angle appears smaller in size when adopting circular radii as size measurement instead of semi-major axis half-light radii.
We therefore consider the semi-major axis half-light radius and its more pronounced difference in size a robust result independent of inclination.

In the following sections, we aim to explore if this difference is intrinsic to the underlying properties of these two populations or a consequence of either inclination or adopting rest-frame optical light profiles as a potentially biased tracer of size.

\subsection{Evolution of Axis Ratios} \label{subsec:axratio}
\begin{figure}
    \centering
    \includegraphics[width=\linewidth]{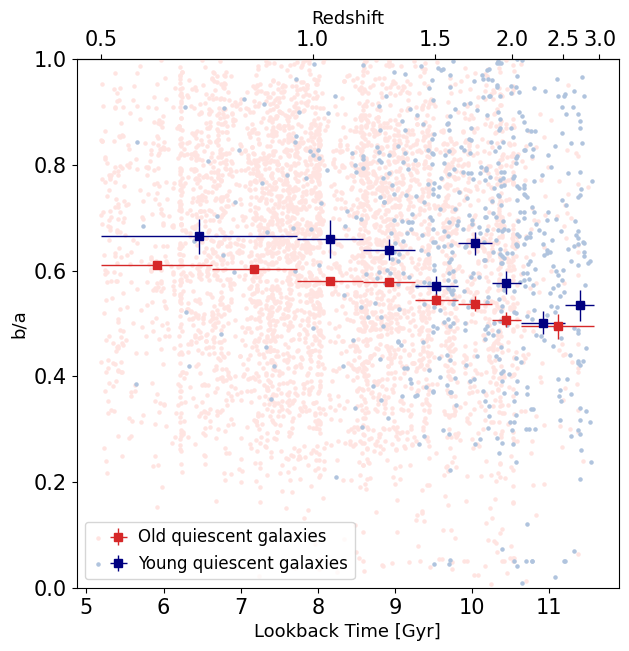}
    \caption{Median axis ratios of old and young quiescent galaxies. The axis ratios of both populations increase toward lower redshift. Interestingly young quiescent galaxies appear less disky than old quiescent galaxies at all redshifts.}
    \label{fig:axratio}
\end{figure}

Next, we explore the evolution of axis ratios.
Generally we think of star-forming galaxies as disky and quiescent galaxies as elliptical \citep[e.g.,][]{vanderwel2009,Patel2013,Chan2021}.  
However, the predicted axis ratios within the quiescent population depend on the formation and quenching pathways.  
If quenching is a gradual process, one might therefore expect the recently quenched galaxies to have lower axis ratios (similar to their star-forming progenitors) relative to old quiescent galaxies.  
The subsequent impact of minor mergers and accretion could puff up these galaxies, with systematically higher (rounder) axis ratios on average. 
However, it may also be that late stage physical mechanisms that ultimately lead to the quenching of these galaxies (e.g., violent disk instabilities or major gas-rich mergers driving gas towards the center) drive their axis ratios higher at this earlier stage.
Figure \ref{fig:axratio} depicts that while the axis ratio of both populations is increasing on average towards lower redshift, the young quiescent sample is at all redshifts larger (rounder) than the old sample on average.

Figure 4b of \citet{Nelson2023} provides a nice demonstration of the distribution of observed axis ratios and example population models with various intrinsic 3D shapes. 
Spheroids typically have axis ratios ranging from q=0.8-1, whereas the peak of the distribution for disks is at q$\sim$0.5 with a wide dispersion caused by the range of possible orientation angles for these objects. 
Axis ratios need to be considered for the population as a whole; here we see a trend towards more disk-like axis ratios at $z>2$ for all quiescent galaxies. 
Young quiescent galaxies tend to be more spheroid-like overall, but the formal average axis ratio is less than a pure spheroid distribution. 

If these young systems were more disky, dependent on the viewing angle, they could appear smaller when adopting circular radii (see Section §\ref{subsec:size}).
Therefore the fact that young quiescent galaxies appear to be on average more spheroid-like cannot be causing the difference in size relative to the older population.
Additionally this would not affect the semi-major half light radius, since it is relatively independent of viewing angle. 
We therefore turn to other properties possibly influencing the size measurements.

\subsection{Impact of Color Gradients}
\label{subsec:colorgradients}
\citet{suess2019a} show that rest-frame optical half-light radii are a biased tracer of mass and therefore may not be a sufficient measure of galaxy size on their own, especially for $z<1.5$.
Our observed difference in size could therefore be a result of radially varying star formation histories within the galaxies.
Accretion of young, bright stars in the outskirts of old quiescent galaxies or remnants of star formation at the center of young quiescent galaxies could make them appear less/more compact, respectively.
Here we search for evidence supporting whether this is in fact the effect that drives the differences in the left panel of Figure \ref{fig:sizeevolution}.

\begin{figure}
    \centering
    \includegraphics[width=\columnwidth]{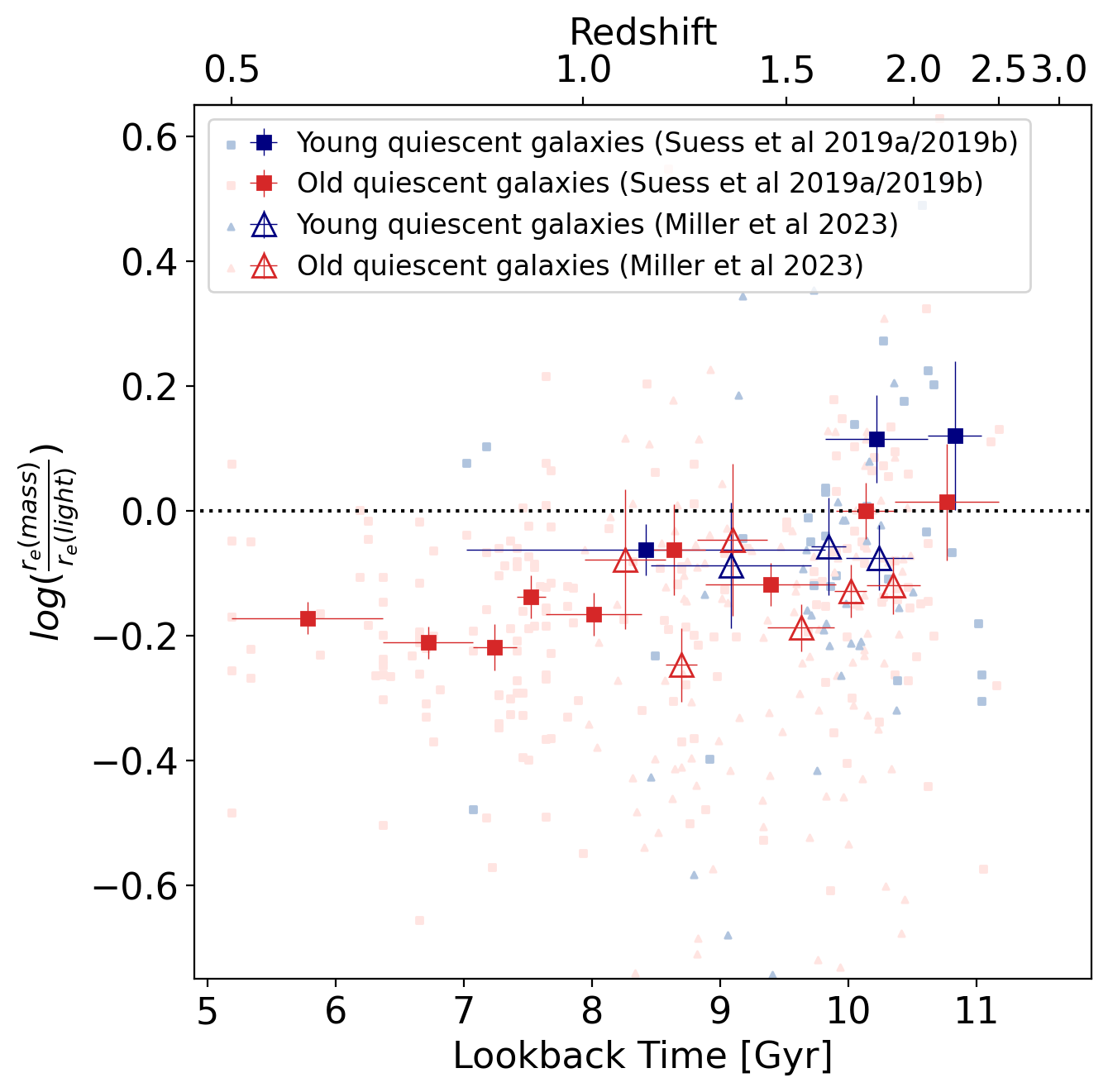}
    \caption{The mass to light ratios of the overlapping galaxies of COSMOS2020 with galaxies from \citet{suess2019a,suess2019b} (squares) and \citet{miller2023} (triangles), respectively. There is no significant difference in the mass to light size ratio between the young and old samples. Both demonstrate negative color gradients at lower redshift, switching to positive color gradients towards earlier cosmic times.}
    \label{fig:colorgradient}
\end{figure} 
In Figure \ref{fig:colorgradient} we show the ratio of effective half-mass to half-light radii for all galaxies in our sample that have measured half-mass radii from either  \citet{suess2019a,suess2019b} or \citet{miller2023}. 
For the derivation of both quantities, we refer the reader to the respective papers.
We defer our own analysis of color gradients to future work based on deeper near-IR imaging (from, e.g., JWST); the measurements of color gradients based on 3D-DASH imaging is not feasible.  
Given this limitation, we instead move forward by evaluating this smaller sample from the literature through the lens of young and old sub-populations.
The black dotted line indicates equality between effective half-mass and half-light radii.
Because the number of overlapping galaxies within our selection criteria is much smaller with this sample -- 181 old and 36 young quiescent galaxies in the work of \citet{suess2019a,suess2019b} and 122 old and 39 young quiescent galaxies in \citet{miller2023} -- we require a minimum of 10 (20) young (old) galaxies per bin to reduce uncertainties resulting from low number statistics. 
These earlier works calculate the mass-weighted sizes adopting independent approaches, but both find that quiescent galaxies have larger light-weighted sizes relative to their mass-weighted sizes on average at $z<1.5$.

For galaxies that overlap with our sample, Figure~\ref{fig:colorgradient} shows that young quiescent galaxies have slightly higher $r_{\rm{e,mass}}/r_{\rm{e,light}}$ ratios than older galaxies at a given epoch; however, the difference is typically within 1$\sigma$. 
At no point within our selected redshift range is there a significant difference in log$(\frac{r_{e,mass}}{r_{e,light}})$ between the two samples.

However, one (important) variable we have not yet considered in detail in our analysis is stellar mass. We explore the impact of stellar mass in the following sub-section.  See also the discussion in Section \ref{sec:discussion} for a  more detailed interpretation of the impact of color gradients.

\subsection{Impact of Stellar Mass} \label{subsec:massdependence}
  
\begin{figure*}
    \centering
    \includegraphics[width=\textwidth]{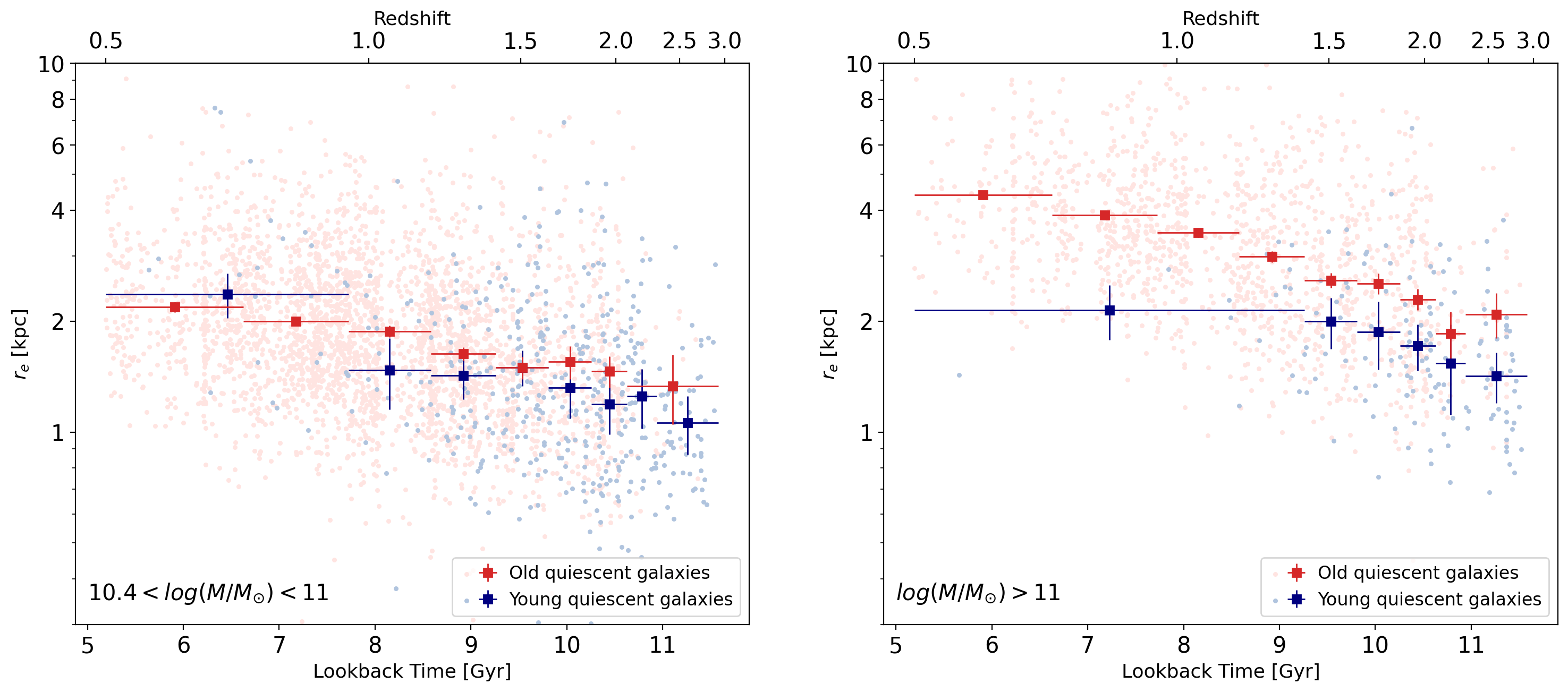}
    \caption{Evolution of effective half-light radii for two sub-populations of 10.4$<$log(M$_{\star}$/M$_\odot$)$<$11 (\textbf{left panel}) and $>$10$^{11}$ M$_{\odot}$ (\textbf{right panel}) respectively. The size difference between the old and young quiescent galaxies disappears when only looking at intermediate mass galaxies, whereas the size difference becomes even more prominent in the most massive galaxies. }
    \label{fig:sizeevolution_massdependent}
\end{figure*}

\citet{Suess2020} demonstrate that the color gradients and resulting ratios of mass-to-light weighted sizes are stellar mass dependent: while the discrepancy between the sizes of young and old galaxies is effectively removed on average when accounting for color gradients for log(M$_{\star}$/M$_{\odot}$)$<$10.5 and $>$11 (albeit the sample size is quite small on the massive end), there still exists tension at log(M$_{\star}$/M$_{\odot}$)$\sim$10.5-10.8 \citep[see Figure 2b in][]{Suess2020}. 
Given these findings and owing to the fact that we have a larger sample size, we are in a good position to test for the impact on the relative sizes between young and old galaxies when splitting our sample into two stellar mass bins. We thus define a massive sample log(M$_{\star}$/M$_\odot$)$>$11 and an intermediate mass sample 10.4$<$log(M$_{\star}$/M$_\odot$)$<$11.  
We show the resulting size evolution of the sub-populations in each mass bin in Figure \ref{fig:sizeevolution_massdependent}. 

\begin{figure*}
    \centering
    \includegraphics[width=\textwidth]{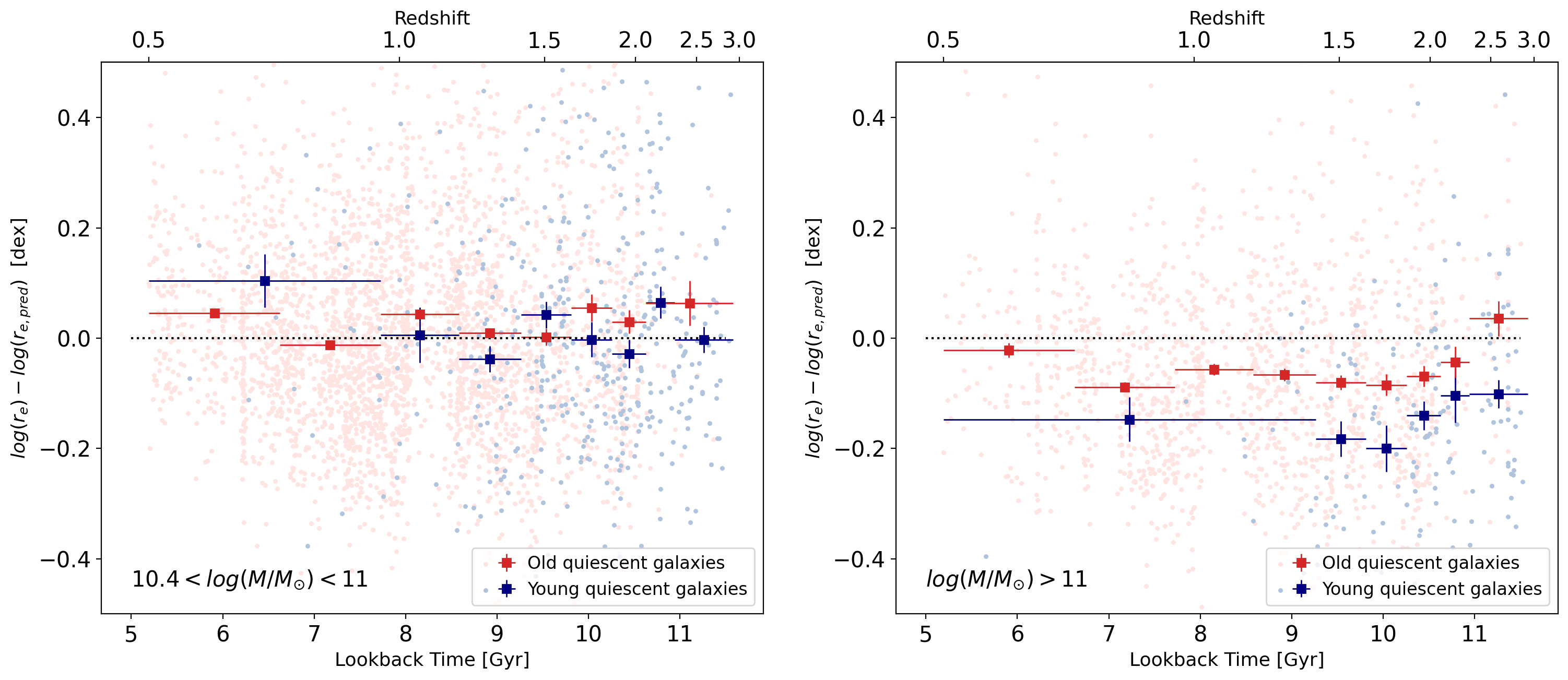}
    \caption{Evolution of the deviation of the measured effective half-light radii from the predicted sizes for two sub-populations of 10.4$<$log(M$_{\star}$/M$_\odot$)$<$11 (\textbf{left panel}) and M$_*>$10$^{11}$ M$_{\odot}$ (\textbf{right panel}) respectively. Again, the intermediate-mass quiescent galaxies following the scaling relations on average, while there is a significant offset towards smaller sizes for the massive old and especially the young quiescent galaxies.}
    \label{fig:sizemassevolution_massdependent}
\end{figure*}
In Figure~\ref{fig:sizeevolution_massdependent}, we demonstrate that the size difference between the two sub-populations is largely driven by the massive sub-sample and less significant or even not present in the intermediate mass sub-populations.  The latter finding is in contrast with the light-weighted sizes of intermediate mass galaxies presented in \citet{Suess2020}.  It may be that the pronounced difference between young and old galaxy sizes observed in this earlier work at intermediate masses was a consequence of a smaller sample size.  Here we are considering a significantly larger sample, with a factor of $\sim$20$\times$ more galaxies, and we show that the size offset is clearly driven by the most massive galaxies that were not well represented previously. 

In order to demonstrate that our incompleteness for intermediate mass galaxies at high redshift does not impact our conclusions, we turn to Figure \ref{fig:sizemassevolution_massdependent}. Here, we show the deviations from the size-mass relation of the intermediate mass and massive sub-samples of old and young quiescent galaxies separately.
The deviations are calculated as the difference between the measured half-light effective radius and the radius predicted by the size-mass relation in log-space (see Equation \ref{eq:arjen_re}).
While Equation \ref{eq:arjen_re} seems to describe the intermediate mass quiescent galaxies of both populations reasonably well (as seen in the left panel), the size-mass relation for massive quiescent galaxies appears to be more complex.
The right panel of Figure \ref{fig:sizemassevolution_massdependent} clearly shows that the radii of young quiescent galaxies are significantly smaller at all redshifts when assuming the same size-mass relation for both populations. The offset from the scaling relations (indicated by the dotted line) for old galaxies is an artifact of the single size-mass relation adopted here.

Other studies assume a broken power-law \citep[e.g.,][]{mowla2019,Cutler2022}, with the pivot mass close to the adopted lower mass limit of this study.  The result is a shallower relation on average at the massive end, which would account for the overall offset among the quiescent population.  Regardless, the important point here is that the \emph{relative} offset remains at all redshifts considered.

\subsection{Impact of Surface Density}
\begin{figure}
    \centering
    \includegraphics[width=\columnwidth]{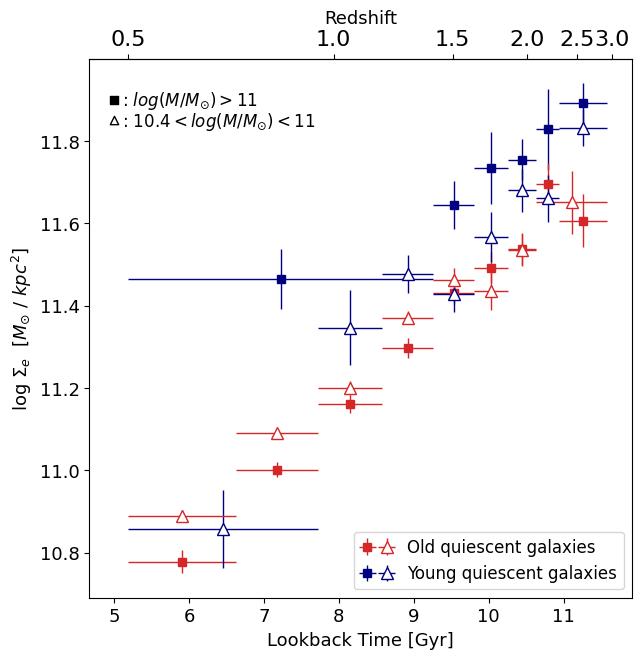}
    \caption{Evolution of the stellar surface density within the half-light radius ($\Sigma_e = M_*  / (2\pi r_e^2$)) for two sub-populations of 10.4$<$log(M$_{\star}$/M$_\odot$)$<$11 (open triangles) and M$_*>$10$^{11}$ M$_{\odot}$ (filled squares) respectively. There is a significant offset in surface density at all redshifts between intermediate mass and massive young quiescent galaxies. Whereas, the average stellar surface densities of old quiescent galaxies agree at the 1$\sigma$ among the massive and intermediate mass older population.}
    \label{fig:compactness_massdependent}
\end{figure}

The previous section demonstrates that the size difference is driven by the most massive galaxies. 
To explore this impact of the stellar mass further, we turn to comparing the surface density of these galaxies dependent on there stellar mass and ages.
To do so, we assume the surface area that includes half of the emitted light to be $\pi r_e^2$, and divide this area by 50\% of the measured stellar mass (i.e., the stellar mass enclosed within the effective radius). Figure \ref{fig:compactness_massdependent} shows that the rate of increase of the average stellar surface density of quiescent galaxies towards earlier times is independent of their age or mass.  These results agree well with e.g., \citet[][]{Barro2017} and \citet{Ji2024}. 

Interestingly, the average surface density of old quiescent galaxies only depends on cosmic time.
Whereas, there is a significant difference in average surface density between the intermediate mass (10.4$<$log(M$_{\star}$/M$_\odot$)$<$11, open triangles) and massive (M$_*>$10$^{11}$ M$_{\odot}$, filled squares) younger sub-populations.
While this difference is more pronounced in the averages of old quiescent galaxies, it is still significant at high redshift in the younger sample.
Similar to the evolution of half-light radii, this general decrease towards lower redshift can be explained by rapid quenching of very massive galaxies that formed relatively early.
Recent spectroscopic studies have confirmed the existence of such systems with quenching times as early as z$\approx$6 \citep[e.g.,][]{Glazebrook,Carnall2023,AntwiDanso2023,Alberts2023,Kakimoto2024}.
Subsequently, these galaxies would then grow through minor mergers, bulking out the outskirts of the galaxies, which decreases the stellar surface density towards later cosmic times \citep[e.g.,][]{Ceverino2015}.
Again, similar to the evolution of half-light radii, Figure \ref{fig:compactness_massdependent} indicates no significant difference between the surface densities of intermediate mass young quiescent galaxies relative to the older intermediate mass population.
However, when considering only the most massive galaxies, the opposite is the case.
\citet{Barro2017} and \citet{Waterval2024} demonstrate that star-forming galaxies have lower stellar surface densities than quiescent galaxies at all redshifts and stellar masses. 
Given this work, we therefore would expect that recently quenched galaxies have lower surface densities than the older population.
However, \citet{Barro2017} focus on $log(M_*)$$<$11 $M_{\odot}$ galaxies and there appears to be considerable overlap between star-forming and quiescent galaxies at what we consider intermediate mass galaxies. 
Since \citet{Waterval2024} focus on the more massive end the tension with our findings are unlikely to be a result of different sample selections. 
We therefore follow the suggestion of \citet{Barro2013}, and posit that the difference in size and surface density could indicate different quenching mechanisms depending on mass.

\subsection{Evolution of Intrinsic Scatter in Size Measurements}
\label{subsec:scatter}
\begin{figure}
    \centering
    \includegraphics[width=\columnwidth]{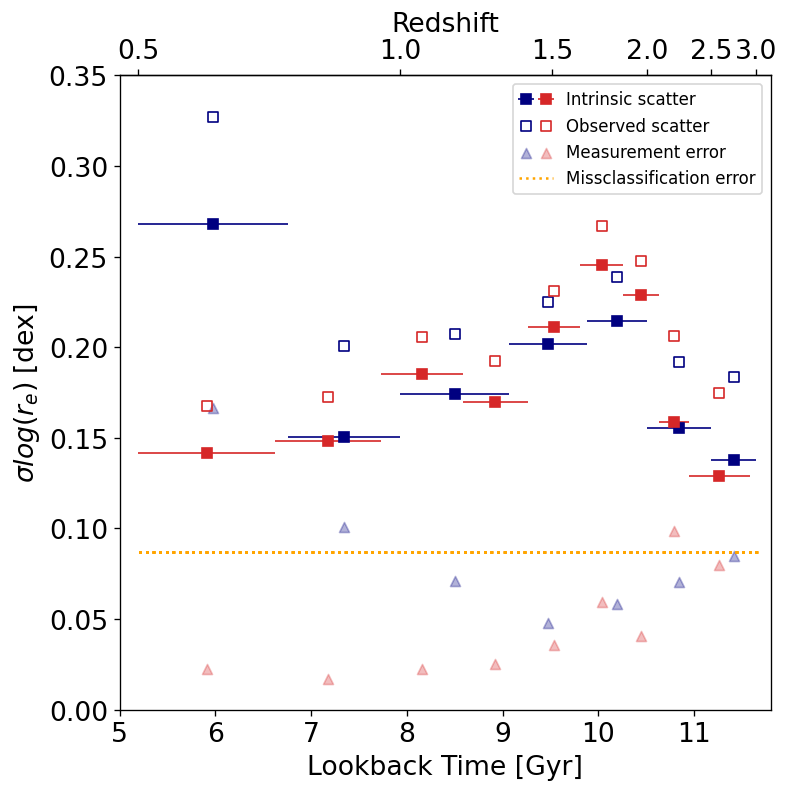}
    \caption{There is a strong co-evolution in intrinsic scatter for young (blue) and old (red) quiescent galaxies, rapidly increasing over $\sim$2 Gyr of cosmic time from $z$=3 to $z$=1.5, and decreasing thereafter. The intrinsic scatter diverges at $z<1$, with the scatter of old quiescent galaxies continuing to decline during the epoch where the population itself is already well established (see Figure~\ref{fig:numberdensity}) and hence simply passively evolving, whereas the intrinsic scatter in the sizes of recently quenched galaxies dramatically increases. The small triangles depict the bootstrapped error while the orange dotted line depicts the assumed missclassification error.}
    \label{fig:errorevolution}
\end{figure}

To better understand the formation pathways of young and old quiescent galaxies, we turn finally to the evolution of the intrinsic scatter in size of these two populations (see Figure \ref{fig:errorevolution}).
A smaller intrinsic scatter would indicate a more uniform distribution in size whereas larger intrinsic variations would result from a wide range of sizes and could imply a broader diversity in formation and quenching pathways, either in terms of the seed progenitor population and/or due to more varied physical processes acting on this population. A standard baseline for the intrinsic scatter is presented in \citet{VanderWel2014}, showing a 0.2 dex intrinsic scatter at all redshifts, but here we can break down the quiescent population further by age.  

Figure \ref{fig:errorevolution} presents the redshift evolution of the observed (open symbols) and intrinsic (filled symbols) scatter of young (blue) and old (red) quiescent galaxies from $z$=3 to $z$=0.5.  We calculate the intrinsic scatter based on the assumption that the observed scatter of the distribution $\sigma_{\mathrm{obs}}$ is the quadratic sum of the intrinsic scatter $\sigma_{\mathrm{int}}$, a measurement error $\sigma_{\mathrm{meas}}$, and a misclassification error $\sigma_{\mathrm{mis}}$.

The measurement scatter is calculated by using a bootstrapping technique as follows. 
We first randomly redraw a new sample from the original sample and perturb the original effective half-light radii measurements within the measurement errors assuming the errors are Gaussian in logarithmic space.
Following \citet{VanderWel2014}, the individual measurement uncertainties includes the formal GALFIT uncertainty added in quadrature to a second term that scales with SNR as follows, 
\begin{equation}
\sigma_{i} = \sqrt{\sigma_{i,\mathrm{galfit}}^2+\left(0.1\left(\frac{\mathrm{SNR_{i,\mathrm{F160W}}}}{50}\right)^{-0.5}\right)^2}
\end{equation}

To include the uncertainty of the stellar mass measurements, we also perturb the stellar mass assuming an uncertainty of 0.15 dex following \citet{VanderWel2014}.
Next we use the perturbed stellar masses and measured redshifts to calculate the expected effective radius assuming a power-law distribution \citep{VanderWel2014}: 
\begin{equation}\label{eq:arjen_re}
    r_e = 10^{A(z)} (M_*/(7\times 10^{10} M_{\odot}))^{\alpha}  
\end{equation}
where $A(z) = -1.57 \times log_{10}(1+z)+0.78$.
We subtract this from the perturbed distribution to account for the shape of the size-mass relation.
We calculate the standard deviation of the difference between the perturbed sizes and those calculated with the perturbed mass measurements based on Equation \ref{eq:arjen_re}.
Repeating this process 100 times leads to a distribution of standard deviations that can be described by a Gaussian with the mean being the original standard deviation of the original sample ($\sigma_{\mathrm{obs}}$) and the width of the Gaussian giving an estimate of the uncertainty of that original standard deviation.
We use this width as an measurement error ($\sigma_{\mathrm{meas}}$) that is then subtracted quadratically together with an estimated miss-classification error of 20\% from the observed scatter of the original (size-mass relation corrected) measurements.  
Note that during this entire process, we handle sizes and errors only in logarithmic space.

We see a marked difference between the intrinsic scatter of old and young quiescent galaxies at $z$$<$1: while the intrinsic scatter in the sizes of young quiescent galaxies abruptly increases at $z\lesssim0.7$, that of old quiescent galaxies instead steadily decreases.  
The number density evolution of the quiescent population tells us that it is largely in place by $z\sim1$ (see Figure~\ref{fig:numberdensity}), with very few young quiescent galaxies added at lower redshifts. 
The increase in scatter ($\sim$0.1 dex larger than the older population) suggests that the few galaxies that do join the quiescent population at later times have a wide range of sizes -- from the larger sizes expected by progenitor bias to the very compact sizes similar to what we observe at high redshift (see also work by \citet{yildrim2017} on local compact massive galaxies).
It should be noted that this spike in intrinsic scatter of the young quiescent galaxies coincides with the known clusters in the COSMOS Field \citep[e.g.,][]{gilli2003,vanzella2005} leading to the assumption of different formation pathways in high density environments.
The diversity of galaxies making up the quiescent galaxy population is indicated by a few exemplary galaxies shown in Figure \ref{fig:sizeevolution}.

At  higher redshift $z>1$, we find that the intrinsic scatter of both young and old quiescent galaxies peaks at $z\sim2$ in our sample, but decreases again towards the highest redshifts probed. 
This brief spike in the intrinsic scatter of quiescent galaxies at $z$$\sim$1.7 may be remnant of the rapid growth of the population within the preceding billion years, as this marks the time period where recently quenched galaxies are most rapidly crossing into this older population.  
If for example there is a wide range in the quenching timescales \citep[e.g, see work by][]{carnall2018,akhshik2023}, there could in principle manifest as a spike in the intrinsic scatter.

Quenching at high redshift may occur on much more rapid timescales, as found by \citet{park2022}, who demonstrate that the fraction of young quiescent galaxies that experience rapid quenching significantly increases with redshift (4\% at $z\approx 1.4$ to 23\% at $z\approx 2.2$).
\citet{park2022} also speculate that all quiescent galaxies at high redshift formed through not only a rapid quenching process but also through a short starburst, likely triggered by dissipative processes resulting in similar progenitors.
This scenario would be consistent with the small intrinsic scatter and compact sizes that we measure herein.  Moreover, rapid timescales with central starbursts is one viable way to explain the difference in sizes between young and old (massive) quiescent galaxies at these redshifts.

\section{Discussion} \label{sec:discussion}

In this paper we investigate the interplay between the build-up of the quiescent population in number and their structural evolution by examining how the number densities and morphological parameters of young and old quiescent galaxies evolve over cosmic time.

The number density of old quiescent galaxies rapidly increases from $z$=3 to $z$=2, leveling off at lower redshifts, whereas young quiescent galaxies are rare at late times but dominate the quiescent galaxy population at $z$$>$1.5 (see Section \ref{fig:numberdensity}).
Comparing with previous studies in Figure \ref{fig:numberdensity_comparisson} \citep{whitaker2012,wild2016,forrest2018,belli2019}, we demonstrate that the normalization of the number density is dependent on the chosen stellar mass limit, whereas the overall shape of the redshift evolution we observe is consistent with these previous studies.
Simple evolutionary models that explain the growth of the old population, as fueled by adding younger quiescent galaxies, would suggest that the number densities of young quiescent galaxies should continue to rise at $z>2$. 
In our study, we instead find that the number density of young quiescent galaxies turns over at $z \approx 2$ and follows the old quiescent galaxies distribution, consistently making up half the quiescent galaxies population.  
Considering the onset of incompleteness in mass at $z\sim2$ stemming from the  3D-DASH magnitude limit, this turnover of the number density at $z>2$ of young quiescent galaxies may not hold true when acquiring deeper data.
\citet{gould2023} predict the fraction of young to old quiescent galaxies steadily increases towards higher redshift (65\% at $3<z<4$ and 86\% at $4<z<5$), based on the full COSMOS2020 catalog.

However, it is also worth noting that recent work by \citet{antwi-danso2022} and \citet{gould2023} hint that the UVJ quiescent selection may already be sub-optimal at $z$=2-3, and is especially not ideal at $z$$>$3.  
Additional ambiguity in our color selection may come from slowly-quenching massive quiescent galaxies that never undergo a rapid starburst phase \citep[transitioning via the green valley in lieu of the bluer post-starburst color-space; e.g.][]{Barro2013,Schawinski2014,carnall2018,belli2019}.  
The consequence of slow quenching is that the bimodality between red and blue galaxies in rest-frame color space becomes increasingly ambiguous at higher redshifts, with more `green valley' galaxies existing. 
These slower quenching galaxies are not necessarily selected with the UVJ method (as adopted herein), and if included would be considered older within this analysis by virtue of their redder V-J colors.  
Hence our simplistic number density estimates may be biased against this particular formation pathway within the young population. 
Taken together, we caution the reader to not over-interpret this turn over in the number density evolution at the highest redshifts.  At face value, our conclusions regarding the smaller sizes measured for young quiescent galaxies also only applies to more rapidly quenching galaxies by design (but see more discussion below).  
We defer a more thorough analysis of the number density evolution of quiescent galaxies at these higher redshifts to data sets that include deeper imaging with longer wavelength coverage from the James Webb Space Telescope \citep[e.g.,][]{casey2022}.

We demonstrate (see Section §\ref{subsec:massdependence}) that massive ($\mathrm{\log(M_{\star}/M_{\odot})}>11$) young quiescent galaxies are significantly smaller than old quiescent galaxies of similar stellar mass at $z>1$.
In the following discussion, we aim to explore whether this difference in size is in fact intrinsic to the population.

One big caveat of our analysis is the rising incompleteness of our sample beyond $z>2$.
It is natural to assume these missing intermediate mass galaxies at these highest redshifts might influence the median size difference between young and old quiescent galaxies.
However we demonstrate in Figure \ref{fig:sizeevolution_massdependent} that the size difference is primarily driven by the most massive galaxies ($log(M/M_{\odot})>11$).
Comparing to Figure \ref{fig:redshicft_mass} it is evident that this massive sample is complete for both sub-populations.
Therefore we consider the observed size difference to not be a result of incompleteness.

The shallow nature of the 3D-DASH survey introduces the possibility that we are missing low surface brightness galaxies in our analysis.
These galaxies would have larger radii and if preferentially missing from the young quiescent population could explain the observable size difference.
We compare the histograms of surface brightness density of the massive ($\log(M/M_{\odot})>11$) populations that drive the observed size difference to the same property derived for a similarly selected sample of galaxies from the deeper but smaller volume 3D-HST survey \citep[][]{VanderWel2014,Skelton2014}.
While the results indicate that we might be missing some low surface brightness galaxies in the young quiescent massive galaxies sample at higher redshift,  it should be noted that this comparison is significantly impacted by the small number of massive young quiescent galaxies at high redshift in the 3D-HST sample (only 19). Therefore, the impact of low surface brightness incompleteness at high redshift and high masses can only sufficiently be determined by a data set that is larger than 3D-HST and deeper than 3D-DASH.  The tentative findings of this paper stand to be tested with upcoming public JWST data sets to confirm the observed size difference.

It is possible that the observed difference is not a true difference intrinsic to the population but rather the effect of ``outshining''.
If galaxies have relatively young stellar populations and/or significant dust obscuration, either as a whole or in certain spatial regions within the galaxy, the effective half-light radius might not be a good tracer of the true size of these galaxies.
This would be indicated by the presence of a color gradient.
Albeit for a significantly smaller sample (with deeper, higher quality imaging), Figure \ref{fig:colorgradient} demonstrates that this outshining is present in young quiescent galaxies at lower redshifts through strongly negative color gradients. However, outshining becomes less significant at higher redshift.
It is therefore possible that the size difference in the massive sub-population at low redshift is not intrinsic to the population. 
To evaluate dust obscuration, we calculated the mean dust attenuation for the young quiescent sample at 0.39 mag with a scatter of 0.35mag, which is significantly lower, than the 0.50 mag with a scatter of  0.37 mag observed in the old quiescent population. Considering the huge scatter in dust attenuation between galaxies, we conclude that it is unlikely that the trends presented herein are solely caused by the presence of dust.
To further ensure a fully quenched sample we remove any UVJ selected galaxy that is within 0.5 dex of the SFR-$M_*$ main sequence, as determined by \citet[][]{Whitaker_2014}, and recover the trends presented herein.
We note, however, that the photometry adopted effectively only gives us information about the stellar populations within the effective radius, and thus with the present analysis we cannot rule out ongoing star formation in the outskirts of these galaxies.

While it is beyond the scope of this paper to perform a radial color gradient analysis for the DASH sample (both due to the shallow depth, but also the lack of longer wavelength coverage), adding high resolution 4.4 micron imaging from JWST is demonstrated to better trace the mass-weighted size of galaxies (REFS), and therefore is the best test to indicate if the light weighted sizes are biased tracers.  Early work with JWST by \citet[][]{Suess2022} finds that F444W sizes are about 10\% smaller on average relative to F150W (similar wavelength to this study); overall the effect is subtle.
A future study will properly test for the effects of gradients in dust, age, and star formation rate, whereas we focus herein on the interpretation of the global morphological properties.

On the other hand, young massive quiescent galaxies at higher redshift may truly be smaller than their older counterparts.
This concurs with both \cite{miller2023} and \citet{vanderwel2023}.
The data in Figure \ref{fig:colorgradient} are derived from \citet{suess2019a,suess2019b} and \citet{miller2023}. 
However both samples do not include enough massive ($\log(M/M_{\odot})>11$) young quiescent galaxies to definitively rule out the assumption of outshining driving the evident size difference in this sub-population.
While the large area of 3D-DASH allows for the detection of an unprecedented large sample of recently quenched, massive galaxies, the shallow nature of the imaging significantly limits our ability to derive meaningful half-mass radii through color gradient analyses.  
Such an analysis is well within the capabilities of large JWST imaging surveys, given the large telescope collecting area and the moderately wide NIRCam field of view, and will be the topic of work. 

If we take the fact that massive young quiescent galaxies are on average smaller than old quiescent galaxies at $z>1$ with a small intrinsic scatter at face value, these results are congruent with a model where quenching is moderately uniform at fixed stellar mass at high redshift.  
One favored theory assumes that galaxies at high redshift quench rapidly after a brief burst of star-formation, driven either by violent disk instabilities \citep[e.g.,][]{Ceverino2015} or gas-rich mergers \citep[e.g.,][]{wellons2015}.  
Their subsequent size growth over the next $\sim$10 billion years is dominated by accretion and minor mergers \citep[e.g.,][]{newman12}.

Given that the the difference in average  half-light radii as well as surface density is driven by the most massive galaxies, quenching mechanisms might depend on the total stellar mass of the galaxies at high redshift \citep[e.g.,][]{Barro2017, Ji2024}.
Whereas intermediate mass galaxies follow a less abrupt and possibly slower pathway towards quiescence.

At more recent times other quenching processes may be present.
One of which could include a much slower seizing of star-formation controlled by the environment \citep[e.g.,][]{peng2010}.  
The consequence of a diverse range of formation pathways is a larger scatter in the (moderately small in number) recently quenched population that would preferentially occur at later times in the context of hierarchical structure formation models. At lower redshift, we also probe a wider range of large-scale environments such as galaxy clusters \citep[e.g.,][]{gilli2003,vanzella2005}, which we speculate contributes to the increased scatter of young quiescent galaxies at $z<1$ measured herein. Once galaxies quench, they are also observed to grow via dry mergers \citep[e.g.,][]{vandokkum2010,newman12,belli15}. This is a slow process acting in tandem with more commonly occurring minor mergers and accretion that slowly puff up the outer envelopes of quiescent galaxies \citep[e.g.,][]{hill2016}.  The result is the average size of quiescent galaxies increases with time and once outshining wears off, the intrinsic scatter slowly becomes smaller, and the population becomes more uniform with time.

\section{Summary and Conclusion} \label{sec:conclusion}

In this paper we investigate the evolution in number density and size of a sample of young and old quiescent galaxies between the redshifts of $0.5<z<3$. Our sample of massive (log(M$_{\star}$/M$_\odot$)$>$10.4) quiescent galaxies is selected from the COSMOS2020 photometric catalogs \citep{weaver2022}, with structural measurements made on HST/F160W imaging from the 3D-DASH survey \citep{Mowla2022}.
This study effectively pulls the studies in the literature together into a more cohesive story, highlighting discrepancies that stem from comparing different redshift epochs and/or stellar mass samples rather than real physical discrepancies. 
We find the following:
\begin{itemize}
    \item Young quiescent galaxies increase in number density towards earlier cosmic times. Young quiescent galaxies make up at least half the population of quiescent galaxies at $2<z<3$.
    \item Young quiescent galaxies are on average smaller than old quiescent galaxies at high redshift ($z \gtrsim 1$), but have the same average size at $0.5<z<1$.
    \item Significant deviations from the size-mass relation exist when additionally considering the age of the population. In particular, the smaller size of young quiescent galaxies compared to old quiescent galaxies at high redshift is driven by the more massive population (M$_*>$10$^{11}$ M$_{\odot}$), with no significant difference measured between the sizes of young and old intermediate mass quiescent galaxies.
    \item More massive (M$_*>$10$^{11}$ M$_{\odot}$) quiescent galaxies exhibit lower stellar surface densities than intermediate mass quiescent galaxies. Furthermore, young quiescent galaxies are significant more compact than their older counterparts at high masses. Interestingly, this is not the case for intermediate mass galaxies, with no statistically significant difference between the average stellar surface densities of young and old quiescent galaxies.
    \item The intrinsic scatter in the sizes of the young and old quiescent population shows a similar redshift evolution from $z=3$ to $z=1$, with a peak coinciding with period at $z=1.5-2$ where half of quiescent galaxies are young. However, the scatter is 0.1 dex larger among young quiescent galaxies at $z<1$. This implies that galaxies that quench at later cosmic times do so through a diverse range of pathways, whereas the physical mechanisms driving quenching at higher redshift most commonly yield compact remnants.
\end{itemize}

While outshining may explain any offsets in size among young and old quiescent galaxies at $z<1.5$, the difference appears to be real at higher redshifts.  
The launch of JWST has allowed further study of the mass-to-light ratio \citep[e.g.,][]{miller2023,vanderwel2023,Suess2022} due to its longer wavelength range extending to 4$\mu$m.
With this work we emphasize the need for multi-wavelength observations of these rare massive young quiescent galaxies at high redshift.
The 6.5m diameter mirror of JWST as well as the larger field of view of NIRCam allows for much more efficient mapping of large areas to hopefully allow for a formal color gradient analysis.  
We therefore look forward to future work that will help us understand the deviation in size between young and old quiescent galaxies which might result in new insights into the mechanisms causing quiescence.
We anticipate these future studies to further illuminate the formation and quenching pathways driving the trends measured herein.  
\section*{Acknowledgement}

The authors thank the anonymous referee for a careful read of the paper and useful comments.
This research is based in part on observations made with the NASA/ESA \emph{Hubble Space Telescope} obtained from the Space Telescope Science Institute, which is operated by the Association of Universities for Research in Astronomy, Inc., under NASA contract NAS 5–26555. These observations are associated with programs HST-GO-09822, HST-GO-14114 and HST-GO-16259.  All of the data presented in this paper were obtained from the Mikulski Archive for Space Telescopes (MAST) at the Space Telescope Science Institute. The specific observations analyzed can be accessed via \dataset[DOI: 10.17909/srcz-2b67]{https://doi.org/10.17909/srcz-2b67}. KEW gratefully acknowledges funding from HST-GO-16259 and the Alfred P. Sloan Foundation Grant FG-2019-12514.  The Cosmic Dawn Center is funded by the Danish National Research Foundation (DNRF) under grant \#140.

\bibliography{sample631}{}
\bibliographystyle{aasjournal}

\end{document}